\documentclass{iaus}
\usepackage{graphicx}

\title[Virgo Early-Type Dwarfs in ALFALFA] 
{Virgo Early-Type Dwarfs in ALFALFA}

\author[Rebecca A. Koopmann]   
{Rebecca A. Koopmann$^{1,2}$%
 }

\affiliation{$^1$ Department of Physics and Astronomy, Union College,
  Schenectady, NY 12308, \break email: koopmanr@union.edu \\
$^2$ National Astronomy and Ionosphere Center\thanks{
The National Astronomy and Ionosphere Center 
is operated by Cornell University under
a cooperative agreement with the National Science Foundation.}, Space Sciences
   Building, Cornell University, Ithaca NY 14853}

\pubyear{2007}
\volume{244}  
\pagerange{xxx-xxx}
\date{?? and in revised form ??}
\jname{Proceedings Title IAU Symposium}
\editors{Jonathan I. Davies \& Michael D. Disney, eds.}

\def\etal{{\it et al.}}

\def\msun{$M_\odot$}

\begin{document}

\maketitle

\begin{abstract}
Early-type dwarf galaxies dominate cluster populations, but their
formation and evolutionary histories are poorly understood.  The
ALFALFA (Arecibo Legacy Fast ALFA) survey has completed observations
of the Virgo Cluster in the declination range of 6 - 16 degrees.  Less
than 2\% of the early-type dwarf population is detected, a
significantly lower fraction than reported in previous papers based on
more limited samples. In contrast $\sim$30\% of the irregular/BCD dwarf
population is detected.   The
detected early-type galaxies tend to be located in the outer regions of the
cluster, with a concentration in the direction of the M Cloud.  Many
show evidence for ongoing/recent star formation.  Galaxies such as these
may be undergoing morphological transition due to cluster
environmental effects.

\end{abstract}

\firstsection 
\section{Introduction}

Early-type dwarfs are often modeled as forming their stars in a single
burst early in the history of the Universe.  Yet some early-type
dwarfs show signs of relatively recent star formation, as revealed by
stellar populations, star formation, and gas content (e.g., Conselice
\etal~2003; Lisker \etal~2006).  The detection of recent star
formation in some systems suggest that at least some early-type dwarfs
have formed by transitioning from another galaxy class, e.g., a
later-type spiral or dwarf that loses its gas as it enters the hostile
environment of a galaxy cluster.  Lisker \etal~(2007) find that
approximately half of the early-type dwarfs in Virgo belong to an
unrelaxed population that could be associated with recently infalling
galaxies.

Conselice \etal~(2003) suggested that up to 15\% of Virgo Cluster dE
galaxies have an HI reservoir. However this study was based on
observations of a total of only 48 objects and used results from a
heterogeneous set of telescopes.  The Arecibo Legacy Fast ALFA
(ALFALFA) Survey, a sensitive blind survey of the Arecibo sky
(Giovanelli \etal~2005 and these proceedings), 
is providing a complete and unbiased view of
HI content and structures in the entire Virgo cluster region.

\section{Observations and Results}\label{sec:concl}

\begin{figure}
\includegraphics[scale=0.35]{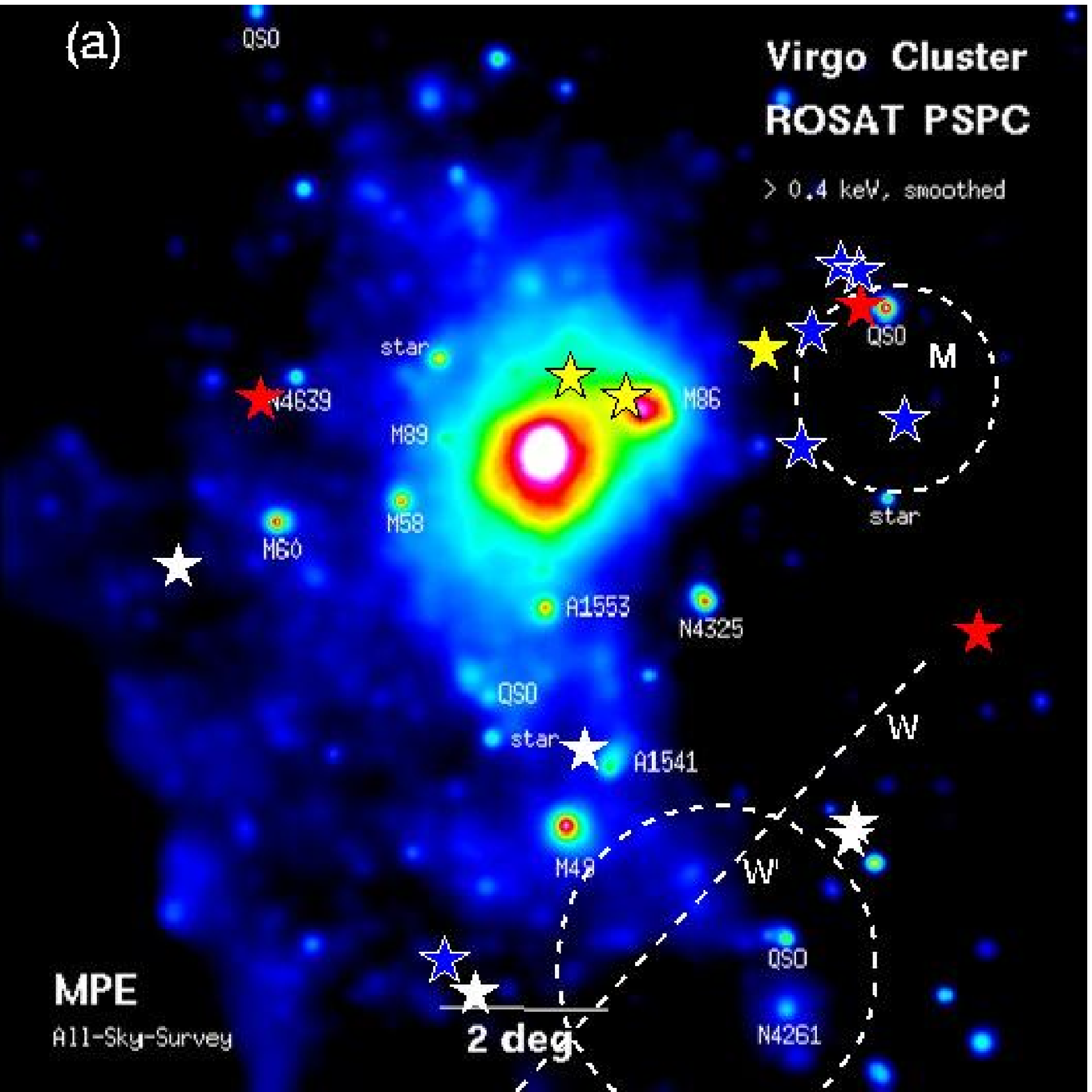}
\hspace{-0.1in}
\includegraphics[scale=0.42]{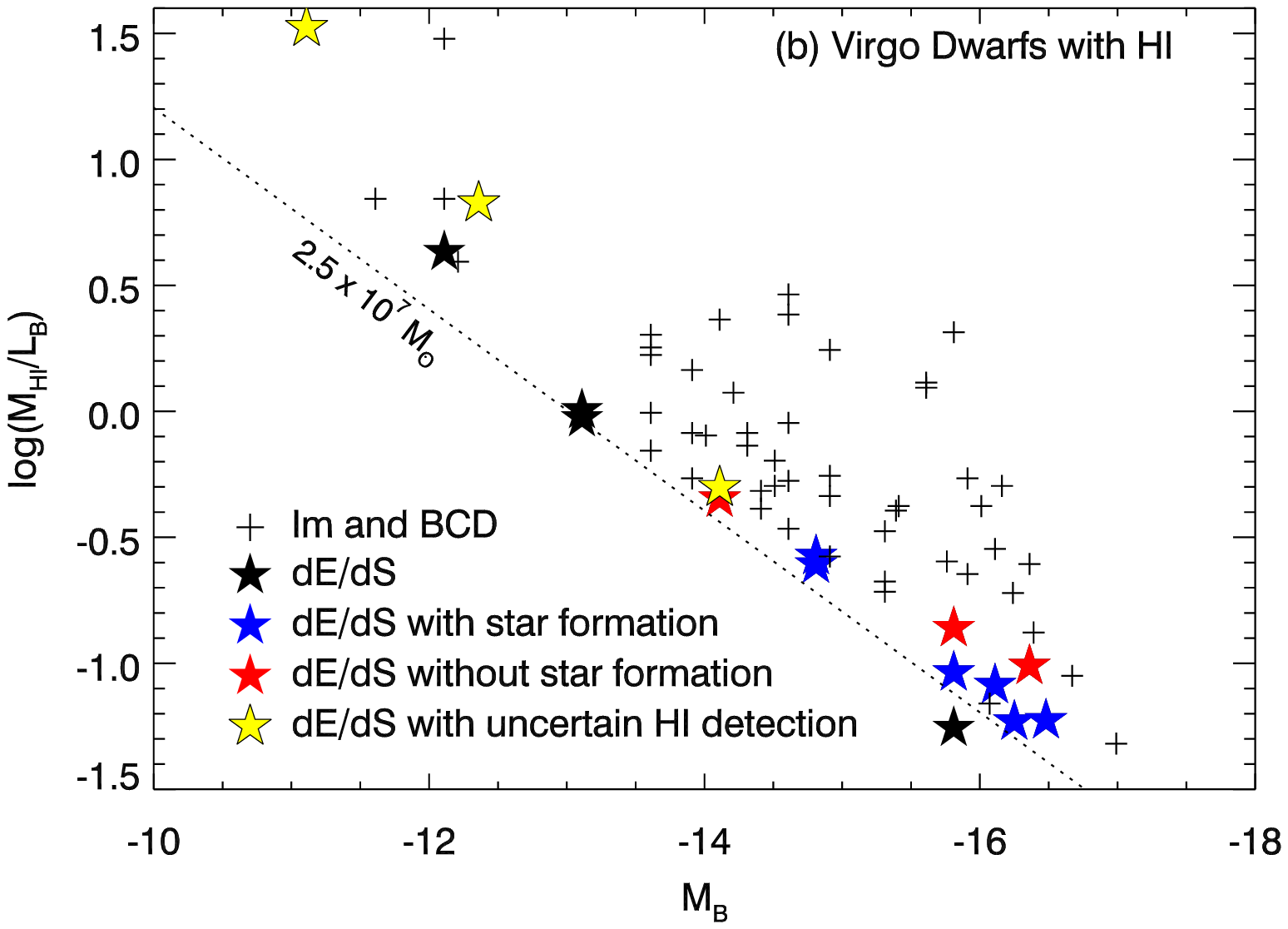}
\caption{(a) Positions of Virgo early-type dwarfs overlaid on 
ROSAT greyscale (Boehringer \etal~1994), with M, W, and W' cloud
positions after Binggeli \etal~(1993).  (b) $M_{HI}/L_B$ as a function
of $M_B$ for Virgo dwarfs with HI detected by ALFALFA. A key to the symbols
is given in the legend.\label{fig1}}
\end{figure}

ALFALFA detections in the Virgo Cluster between the
declinations of 6 - 16 degrees (Giovanelli \etal~ 2007; Kent \etal~
2007, in prep., Koopmann \etal~2007, in prep.) were compared with the
Virgo Cluster Catalog (VCC: Binggeli, Sandage, \& Tammann 1985;
Binggeli, Popescu, \& Tammann 1993) to identify galaxies classified as
early-type dwarfs.  A total of 16 or 1.4\% of the dE/dS0 in this
declination range were detected by ALFALFA. This compares to a
detection rate of 29\% for dwarf irregulars and 27\% for BCDs. 63\% of
the early type detections are new HI measurements and 32\% are new
redshift measurements.  56\% have peculiar or uncertain
classification. Most are located in the cluster outskirts
(Figure~\ref{fig1}a), with several in the direction of the M Cloud. 
Six
of nine observed at H$\alpha$ wavelengths show emission due to star
formation. 

Conselice \etal~(2003) reported 7/48 (15\%) Virgo dE/dS0 sample
detected in HI. Of the five within our declination range, two, the
brightest in their sample, are not detected by ALFALFA and were likely
confused in Conselice \etal~  Three are reproduced by ALFALFA 
with comparable masses, but two of these, VCC 31 (classified \lq ?\rq~ in 
the VCC) and VCC 2062, are of uncertain type and are not counted here as dE/dS0.

The median HI mass of the detected early-types is 3.2 x 10$^7$ \msun~ 
(which is only 1.6 times the ALFALFA detection limit at the 
Virgo distance of 16.7 Mpc) and
the median M$_{HI}$/L$_B$ is 0.14. They have lower
M$_{HI}$/L$_B$ than Virgo dwarfs classified as Im or BCD
(Figure~\ref{fig1}b).  Virgo dwarfs spread to lower
M$_{HI}$/L$_B$ than more isolated dwarfs. 
Detailed results are presented in Koopmann \etal~(2007, in prep.).

\begin{acknowledgments}
The author is grateful for partial support from NAIC, travel support
      from the Mellon Foundation, and for the hospitality of the
      Cornell University Astronomy Department during a sabbatic visit.

\end{acknowledgments}

\end{document}